\documentclass[aps,prb,twocolumn,superscriptaddress,notitlepage]{revtex4-1}
\usepackage{graphicx}
\usepackage[caption=false]{subfig}
\usepackage{float}
\usepackage{natbib}
\usepackage{xcolor}
\usepackage{xr}
\usepackage{amsmath}
\usepackage{amssymb}
\usepackage{array}
\usepackage{wasysym}
\usepackage{color,soul}
\usepackage{braket}
\usepackage{verbatim}
\usepackage{hyperref}
\usepackage{gensymb}
\usepackage{url}
\usepackage{dirtytalk}

\usepackage{filecontents}
\begin{document}

\title{The stripe state at 1/8 Ba doping hosts optimal superconductivity in La-214 cuprates under low in-plane stress}

\author{V. Sazgari}
\affiliation{PSI Center for Neutron and Muon Sciences CNM, 5232 Villigen PSI, Switzerland}

\author{S.S. Islam}
\affiliation{PSI Center for Neutron and Muon Sciences CNM, 5232 Villigen PSI, Switzerland}

\author{M. Lamotte}
\affiliation{PSI Center for Neutron and Muon Sciences CNM, 5232 Villigen PSI, Switzerland}

\author{J.N. Graham}
\affiliation{PSI Center for Neutron and Muon Sciences CNM, 5232 Villigen PSI, Switzerland}

\author{O. Gerguri}
\affiliation{PSI Center for Neutron and Muon Sciences CNM, 5232 Villigen PSI, Switzerland}

\author{P. Král}
\affiliation{PSI Center for Neutron and Muon Sciences CNM, 5232 Villigen PSI, Switzerland}

\author{I.~Maetsu}
\affiliation{Department of Engineering and Applied Sciences, Sophia University, 7-1 Kioi-cho, Chiyoda-ku, Tokyo 102-8554, Japan}

\author{T. Shiroka}
\affiliation{PSI Center for Neutron and Muon Sciences CNM, 5232 Villigen PSI, Switzerland}

\author{G. Simutis}
\affiliation{PSI Center for Neutron and Muon Sciences CNM, 5232 Villigen PSI, Switzerland}

\author{R. Khasanov}
\affiliation{PSI Center for Neutron and Muon Sciences CNM, 5232 Villigen PSI, Switzerland}

\author{R.~Sarkar}
\affiliation{Institute for Solid State and Materials Physics, Technische Universität Dresden, D-01069 Dresden, Germany}

\author{A. Steppke}
\affiliation{PSI Center for Neutron and Muon Sciences CNM, 5232 Villigen PSI, Switzerland}

\author{N.A. Shepelin}
\affiliation{PSI Center for Neutron and Muon Sciences CNM, 5232 Villigen PSI, Switzerland}


\author{M.~M{\"u}ller}
\affiliation{PSI Center for Scientific Computing, Theory and Data, 5232 Villigen PSI, Switzerland}

\author{M. Bartkowiak}
\affiliation{PSI Center for Neutron and Muon Sciences CNM, 5232 Villigen PSI, Switzerland}

\author{M.~Janoschek}
\affiliation{PSI Center for Neutron and Muon Sciences CNM, 5232 Villigen PSI, Switzerland}
\affiliation{Physik-Institut, Universit\"{a}t Z\"{u}rich, Winterthurerstrasse 190, CH-8057 Z\"{u}rich, Switzerland}

\author{J.~Chang}
\affiliation{Physik-Institut, Universit\"{a}t Z\"{u}rich, Winterthurerstrasse 190, CH-8057 Z\"{u}rich, Switzerland}

\author{H.H.~Klauss}
\affiliation{Institute for Solid State and Materials Physics, Technische Universität Dresden, D-01069 Dresden, Germany}

\author{T.~Adachi}
\affiliation{Department of Engineering and Applied Sciences, Sophia University, 7-1 Kioi-cho, Chiyoda-ku, Tokyo 102-8554, Japan}

\author{G.D.~Gu}
\affiliation{Condensed Matter Physics and Materials Science Division, Brookhaven National Laboratory, Upton, New York 11973, USA}

\author{J.M.~Tranquada}
\affiliation{Condensed Matter Physics and Materials Science Division, Brookhaven National Laboratory, Upton, New York 11973, USA}

\author{H. Luetkens}
\affiliation{PSI Center for Neutron and Muon Sciences CNM, 5232 Villigen PSI, Switzerland}

\author{Z. Guguchia}
\email{zurab.guguchia@psi.ch}
\affiliation{PSI Center for Neutron and Muon Sciences CNM, 5232 Villigen PSI, Switzerland}

\date{\today}

\begin{abstract}

The cuprate system La$_{2-x}$Ba$_{x}$CuO$_{4}$ (LBCO) exhibits a pronounced sensitivity to in-plane uniaxial stress, particularly near the 
1/8 doping anomaly, where stripe order strongly suppresses bulk superconductivity.
While previous studies have focused on compositions close to 0.125, the commensurate $x$=0.125 phase remains largely unexplored under symmetry-selective lattice tuning. Here, we combine muon-spin rotation (${\mu}$SR), AC susceptibility, and electrical resistivity to investigate superconductivity, spin-stripe order, and structural response in LBCO-0.125 under in-plane uniaxial stress applied 45$^\circ$ to the Cu–O bond direction. Complementary resistivity measurements on $x$=0.115 and 0.135 track the evolution across both sides of the anomaly. We observe a giant enhancement of the bulk superconducting transition temperature in LBCO-0.125, increasing from 5 K to 37 K under 0.5 GPa. While the onset temperature of spin-stripe order decreases only modestly, the magnetic volume fraction is reduced by about a factor of two, with local order preserved. Simultaneously, the resistivity peak associated with the LTT phase is fully suppressed across all dopings. These results demonstrate that suppression of the LTT phase and reduction of the static spin-stripe–ordered volume fraction are crucial for the development of optimal three-dimensional superconductivity. Strikingly, the composition $x$=0.125, with the most robust stripe stability and the lowest ambient-pressure $T_{\rm c}$ develops the highest
$T_{\rm c}$ under stress, reaching a zero-resistance state at 37 K and an onset of the superconducting transition as high as 46 K. This behavior indicates that stripe-related interactions enhance pairing strength, while static stripe order competes with superconductivity primarily at the level of phase coherence rather than pairing itself. 
A small amount of uniaxial stress is sufficient to suppress the lattice anisotropy that pins stripes, allowing spatially-uniform superconductivity to emerge in an environment of dynamic stripe correlations.

\end{abstract}
\maketitle

\section{Introduction}

The cuprate superconductor 
La$_{2-x}$Ba$_{x}$CuO$_{4}$ (LBCO) serves as a prototypical system that manifests a complex phase diagram arising from the interplay of superconductivity with charge, spin, and orbital degrees of freedom \cite{doi:10.1126/science.237.4819.1133,Tranquada561,Tranquada7489,Emery8814,Keimer179,Guguchia097005,Fradkin457,Huang1161,Vojta699}. Upon cooling, LBCO undergoes a sequence of structural transitions (Fig. 1a-b) from the high-temperature tetragonal (HTT) to the low-temperature orthorhombic (LTO), and finally to the low-temperature tetragonal (LTT) phase \cite{Axe271}. Around the commensurate doping level $x$=1/8, the system stabilizes a stripe-ordered state characterized by spatially modulated spin and charge order. At this doping, static spin and charge stripes are most robust, and three-dimensional superconductivity is strongly suppressed, with the bulk transition temperature reduced to approximately 3 K despite the presence of two-dimensional superconducting correlations at much higher temperatures (Fig. 1c). The stripe order adopts an orthogonal stacking configuration along the 
$c$-axis \cite{Tranquada7489}, which, together with its pair-density-wave character, frustrates the Josephson coupling between adjacent layers. In addition, the centered stacking of stripes in next-nearest-neighbor layers \cite{PhysRevB.107.115125} further suppresses the establishment of three-dimensional superconducting coherence. A central question in cuprate physics is whether superconductivity, magnetism, and charge order compete or cooperate, and how their coupling to lattice distortions governs the emergence of high-$T_{\rm c}$ superconductivity.

\begin{figure*}[!]
    \centering
    \includegraphics[width=1\linewidth]{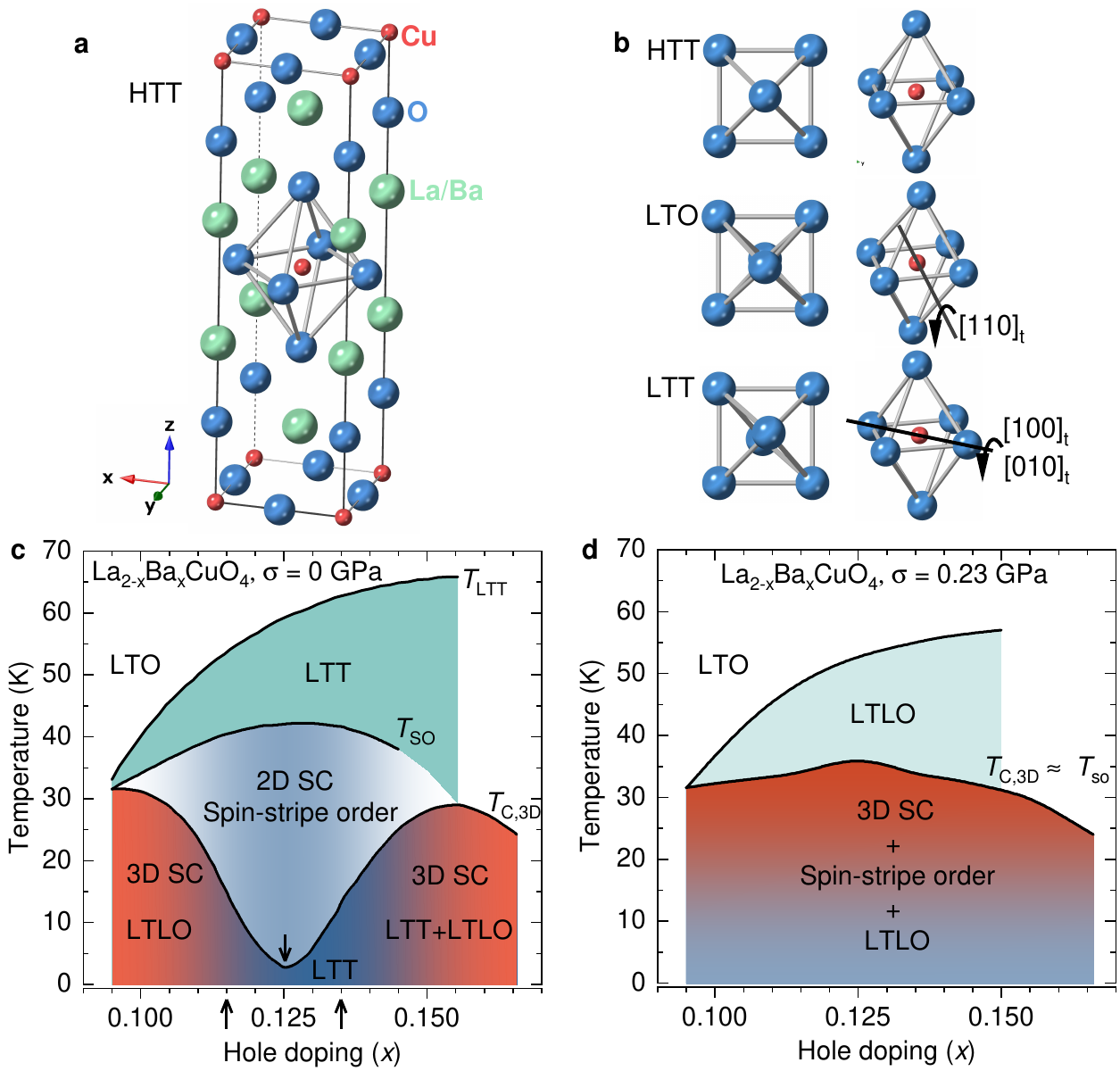}
    \caption{\textbf{Uniaxial Stress Tuning of the LBCO Phase Diagram.} (a-b) Crystal Structure Schematics of LBCO Showing the High-Temperature Tetragonal (HTT), Low-Temperature Orthorhombic (LTO), Low-temperature less Orthorhombic (LTLO)  and Low-Temperature Tetragonal (LTT) Phases. (c-d) Comparison of the Temperature–Doping Phase Diagram of LBCO at Zero Stress and Under 0.23 GPa In-Plane Stress.}
    \label{fig:enter-label}
\end{figure*}

External tuning parameters \cite{Axe2751,Hucker057004,Guguchia093005,Guguchia214511,Axe271,Lieaav7686,Laliberté2018,Chang871} such as hydrostatic pressure, magnetic field and impurity doping have long been used to disentangle these intertwined orders. More recently, in-plane uniaxial stress \cite{Guguchia097005,Guguchiae2303423120,islam2025contrasting, Thomarat271,simutis2022single,Wang1795,BoyleL022004,doi:10.1126/science.aat4708,PhysRevB.108.L121113,doi:10.1143/JPSJ.81.113709,published_papers/46828470,ye2025straininducedstructuralchangenearlycommensurate} has emerged as a particularly powerful and symmetry-selective tuning parameter, capable of dramatically modifying the electronic ground state. Experiments have revealed a strong enhancement of the superconducting transition temperature of LBCO around 1/8-doping under modest in-plane stress \cite{Guguchia097005,Guguchiae2303423120,Thomarat271,https://doi.org/10.1002/adma.202509308}, whereas stress applied along the crystallographic $c$-axis \cite{islam2025contrasting} produces only weak or even negative effects on $T_{\rm c}$. This pronounced anisotropic response highlights the intimate coupling between in-plane lattice symmetry and electronic ordering phenomena.


However, previous studies  on LBCO \cite{Guguchia097005,Guguchiae2303423120,islam2025contrasting,Thomarat271} have primarily focused on compositions near, but not exactly at, the commensurate 1/8 doping—most notably $x$=0.115\cite{Adachi144524} and $x$=0.135.
Moreover, experimental studies at $x$=1/8 have thus far been largely confined to measurements of $T_{\rm c}$ and stress-dependent X-ray diffraction \cite{https://doi.org/10.1002/adma.202509308}. Consequently, the precise $x$=1/8 composition—where static stripe order is maximally stabilized and the suppression of three-dimensional superconductivity is most pronounced—remains insufficiently explored with regard to the microscopic characteristics of the spin-stripe phase (i.e., ordered moment size and magnetically ordered volume fraction) as well as the evolution of the charge stripe order.
In this work, we employ a unique combination of muon-spin rotation (${\mu}$SR), AC susceptibility, and electrical resistivity measurements performed on the same crystal under uniaxial stress conditions, enabling us to simultaneously probe superconductivity, spin-stripe order temperature, spin-stripe ordered volume fraction, and structural signatures across the phase diagram. This approach provides the most comprehensive investigation to date of the stress response in LBCO-1/8. In addition, we present complementary resistivity measurements for dopings $x$=0.115 and $x$=0.135, spanning both sides of the 1/8 anomaly, thereby placing the commensurate composition into a broader electronic context.

These experiments demonstrate that in-plane uniaxial stress unlocks a hidden high-$T_{\rm c}$ state in LBCO-1/8 by suppressing the LTT structural distortion and reducing the static stripe-ordered volume fraction (Fig. 1d). Although stripe order remains long range, the weakening of its static component enables the rapid development of robust three-dimensional superconductivity, with $T_{\rm c}$ rising to 37 K (Fig. 1d). Notably, this transition temperature exceeds the optimal $T_{\rm c}$ achieved for dopings near 1/8 by several kelvin. These findings suggest that stripe-related interactions enhance superconducting pairing, whereas static stripe order primarily suppresses global phase coherence, likely through interlayer frustration. In this framework, uniaxial stress restores interlayer coupling by destabilizing static spin-stripe order and likely promoting dynamical stripe correlations. This, in turn, enables the emergence of a homogeneous $d$-wave superconducting state and shifts the balance from a putative PDW-dominated regime toward a uniform $d$-wave ground state with coherent interlayer pairing.

\section{Results and Discussion}

The diamagnetic response of the LBCO $x$=0.125 crystal, measured prior to mounting in the stress apparatus, is shown in Fig. 2b. The sample was zero-field cooled and subsequently measured in a dc field of ${\mu}_{0}$$H$=0.5 mT. The magnetic field was applied parallel to the CuO$_{2}$ planes, such that the induced shielding currents flow between the layers. This configuration makes the measurement sensitive to the onset of three-dimensional (3D) superconductivity, which occurs near 5 K, consistent with previous reports \cite{Hucker104506}.
To monitor the effect of stress on superconductivity, in situ ac susceptibility measurements were performed at each stress value, either immediately before or after the ${\mu}$SR measurements. The excitation field was applied predominantly along the $c$-axis. The compressive stress was applied at an angle of 45$^{o}$ to the Cu–O bond direction (denoted as [100]) (Fig. 2a). To quantify changes in the superconducting transition temperature, we define the midpoint temperature $T_{\rm c,mid}$, which provides a reliable estimate of the 3D superconducting transition temperature. As shown in Fig. 2b, compressive stress induces a rapid, nearly linear increase of $T_{\rm c,mid}$ from 3 K to 35 K—corresponding to an approximately elevenfold enhancement of 3D superconductivity. The onset of the superconducting transition reaches 37 K. This pronounced enhancement of $T_{\rm c,mid}$ under in-plane stress is qualitatively similar to our previous \cite{Guguchia097005,Guguchiae2303423120} observations for the $x$=0.115 and $x$=0.135 samples. However, several important differences should be emphasized. While the optimal $T_{\rm c,mid}$ achieved under stress in the $x$=0.115 and $x$=0.135 samples was about 30 K, the $x$=0.125 sample reaches $T_{\rm c,mid}$=35K, with an onset as high as 37 K. Moreover, the critical pressure required to reach the optimal superconducting state is approximately three times larger for $x$=0.125. This is consistent with the fact that stripe order is most stable at this doping level. Nevertheless, the critical pressure required to nearly maximize $T_{\rm c,mid}$ remains remarkably small, only about 0.23 GPa. It is noteworthy that even a much larger hydrostatic pressure of 14 GPa increases $T_{\rm c}$ only to approximately 18 K, highlighting the unique effectiveness of uniaxial stress in tuning cuprate superconductivity. An important outcome of these susceptibility measurements is that the $x$=0.125 sample, which hosts the most stable stripe order, ultimately achieves the highest optimal 3D superconducting transition temperature. We essentially reach the optimal $T_{\rm c}$ of LSCO \cite{PhysRevB.40.2254,PhysRevLett.102.047001}, which is highly significant, particularly considering that this occurs at 1/8 doping—where superconductivity is typically strongly suppressed. This conclusion will be further supported by the resistivity measurements presented below.

\begin{figure*}[!]
    \centering
    \includegraphics[width=1\linewidth]{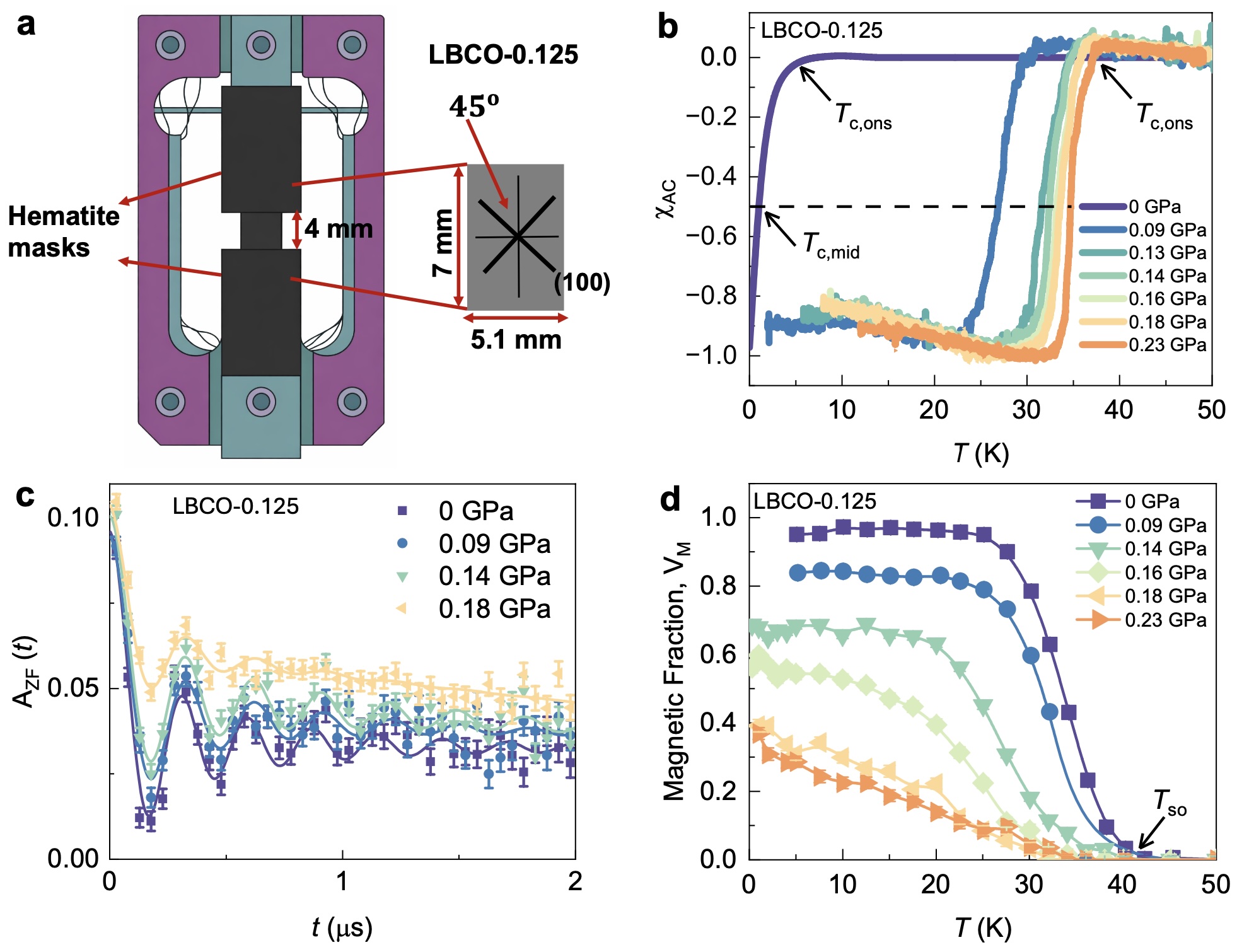}
    \caption{\textbf{Superconducting and spin-stripe response under in-plane stress for LBCO-0.125.} (a) Schematic view of the uniaxial stress holder for ${\mu}$SR experiments with the sample mounted. The view is shown from the direction of the incoming muon beam. The rectangular crystal is glued along the stress axis. The compressive stress was applied at an angle of 45$^{o}$ to the Cu–O bond direction (denoted as [100]). Hematite pieces are positioned to mask the holder-frame regions exposed to the muon beam, thereby reducing background contributions from the pressure cell and ensuring that the ${\mu}$SR signal predominantly originates from the sample. (b) Temperature dependence of the diamagnetic susceptibility measured under various in-plane uniaxial stress conditions up to 0.32 GPa. (c) The zero-field ${\mu}$SR spectra, recorded at the base temperature 700 mK under various stresses. (d) The temperature dependence of the spin-stripe–ordered volume fraction measured under various in-plane uniaxial stress conditions up to 0.32 GPa.}
    \label{fig:enter-label}
\end{figure*}

In the following, we focus on the effect of applied stress on the spin-stripe order. The evolution of spin-stripe order under compressive stress was characterized using a combination of weak transverse-field (TF) and zero-field (ZF) ${\mu}$SR measurements. These experiments allow us to track the temperature and stress dependence of the magnetically ordered volume fraction $V_{\rm M}$ as well as the internal magnetic field at the muon site, which serves as a measure of the ordered moment size. In ZF-${\mu}$SR measurements, the muon spins precess exclusively in the internal magnetic field associated with static magnetic order, with the measured response reflecting an average over the distribution of muon stopping sites relative to the local modulation of the internal field. As shown in Fig. 2c, clear oscillations remain visible under increasing compressive stress, despite a pronounced reduction in signal amplitude. The characteristic internal field $B_{\rm int}$ at the muon stopping site is extracted from the oscillation frequency and exhibits only a slight decrease with increasing stress. This indicates that the static spin/stripe order remains locally robust even at the maximum applied stress. However, the reduction in oscillation amplitude demonstrates a substantial decrease in the magnetically ordered volume fraction. The temperature dependence of magnetic volume fraction $V_{\rm M}$, extracted from weak TF measurements, for various stress values is shown in Fig. 2d. Upon applying stress, the spin-ordering temperature $T_{\rm so}$ decreases moderately, from approximately 40 K at ambient pressure to about 33 K at 0.23 GPa. The internal field exhibits only a modest reduction, and coherent oscillations persist up to the maximum applied stress of 0.23 GPa, indicating the survival of static spin-stripe order. In contrast, the magnetic volume fraction $V_{\rm M}$ decreases much more strongly: at 700 mK, is reduced by approximately a factor of 2.5 at 0.23 GPa. These results demonstrate that although spin-stripe order remains long range with a relatively high ordering temperature, the fraction of the sample hosting static magnetic order is substantially suppressed under stress. This aligns well with our previous observations for the samples with $x$=0.115 \cite{Guguchia097005} and 0.135 \cite{Guguchiae2303423120}, but the characteristic pressure required to reduce $V_{\rm M}$ is higher for $x$=0.125 than for the other two concentrations. 
Specifically, the stress required to reduce $V_{\rm M}$ to 0.4 (40${\%}$) amounts to 0.23 GPa for the $x$=0.125 sample, compared with stresses below 0.1 GPa for the $x$=0.115 and $x$=0.135 compositions. Overall, these are remarkably low pressures, highlighting that even such modest stress is sufficient to substantially perturb the stripe order-even under the most stable conditions.

\begin{figure*}[!]
    \centering
    \includegraphics[width=1\linewidth]{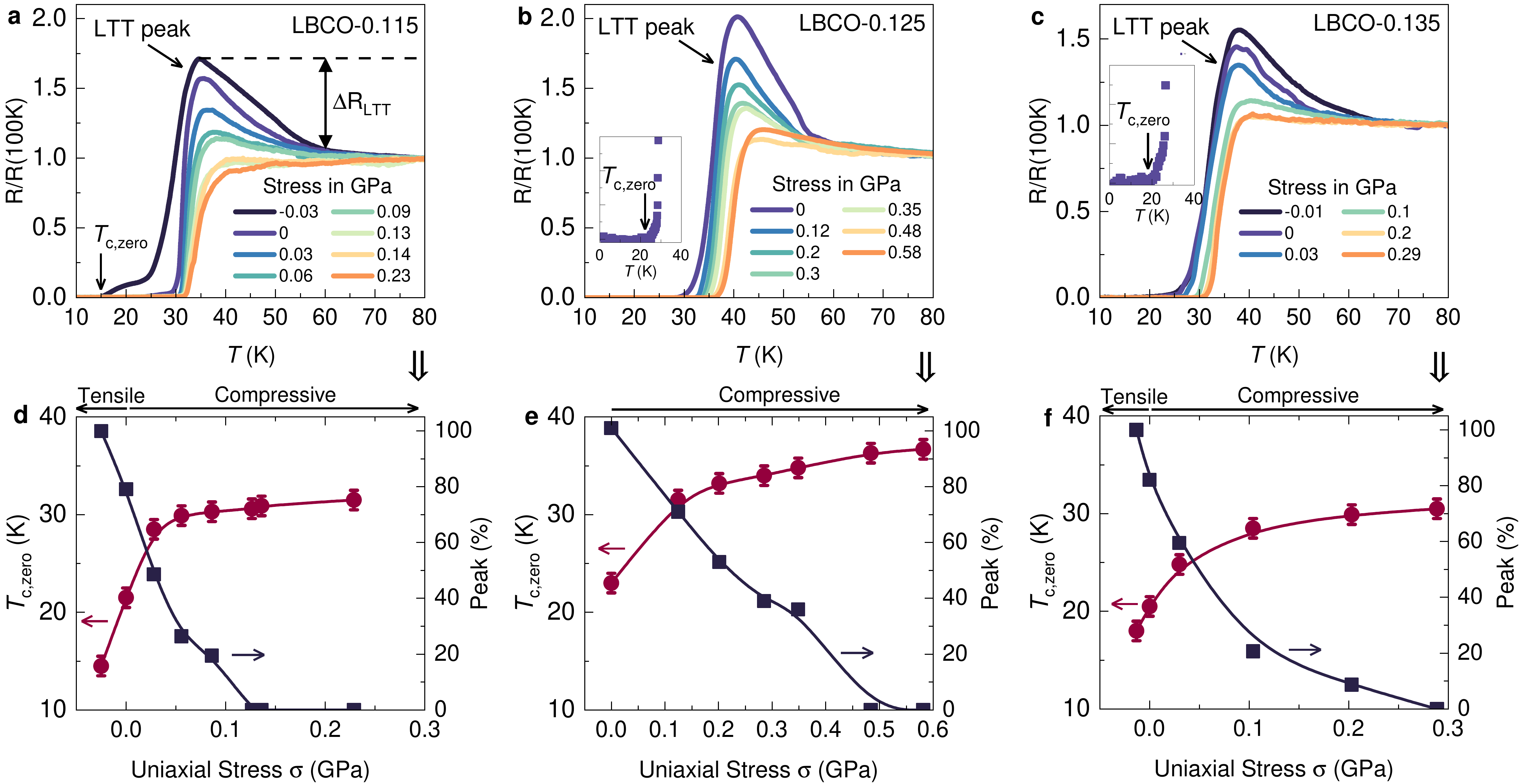}
    \caption{\textbf{Electrical resistivity response of LBCO under in-plane uniaxial stress.} (a–c) Temperature dependence of the in-plane electrical resistance, normalized to its value at 100 K, measured under various in-plane uniaxial stress conditions for three different dopings: $x$=0.115 (a), 0.125 (b), and 0.135 (c). The arrow marks the peak originating from the LTT structural phase transition.(d–f) Stress dependence of the superconducting transition temperature, defined by the onset of zero resistance, and of the normal-state resistivity peak height for the three dopings: $x$=0.115 (d), 0.125 (e), and 0.135 (f). For the $x$=0.125 and $x$=0.135 samples, a small hump just above  $T_{\rm c}$ remains visible even at the highest applied stress, amounting to only a few percent. However, in panels (e) and (f), it was set to zero. }
    \label{fig:enter-label}
\end{figure*}

\begin{figure*}[!]
    \centering
    \includegraphics[width=1\linewidth]{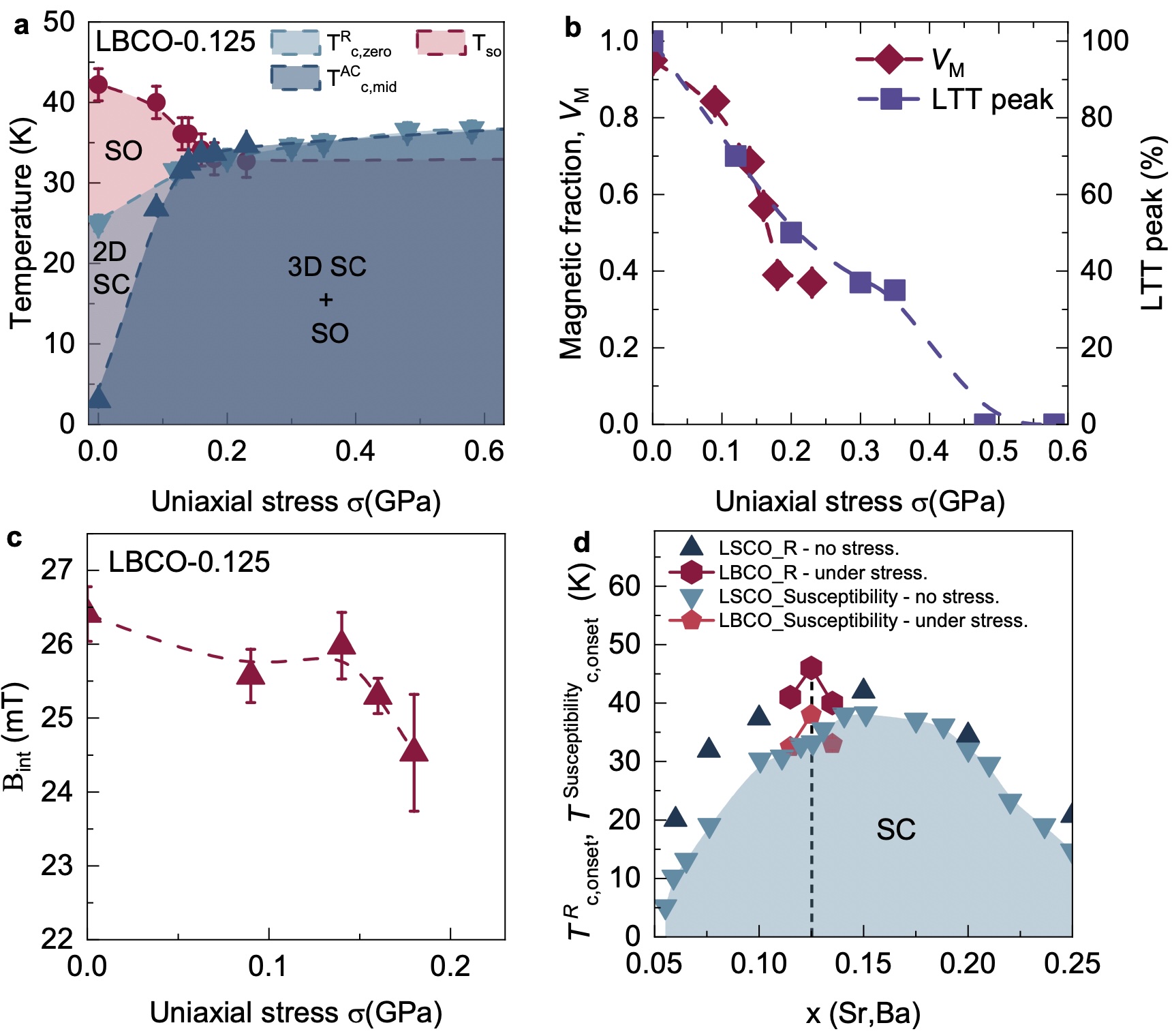}
    \caption{\textbf{Temperature–Stress Phase Diagram for LBCO-0.125.} (a)Stress dependence of the spin-stripe ordering temperature $T_{\rm so}$, the 2D superconducting transition temperature $T^{R}_{\rm c,zero}$ defined by the onset of zero in-plane resistance, and the 3D superconducting transition temperature $T^{AC}_{\rm c,mid}$ determined from the midpoint of the diamagnetic susceptibility transition. (b) Stress dependence of the magnetic volume fraction $V_{\rm M}$ and the normal-state resistivity peak height ${\Delta}R_{\rm LTT}$ associated with the LTT phase. (c) Stress dependence of the internal magnetic field $B_{\rm int}$. (d)  We compare the superconducting transition temperature $T_{\rm c}(x)$ of LSCO at ambient stress with $T_{\rm c}(x)$ of LBCO under applied uniaxial stress tuned to the condition of optimal superconductivity. The onset temperatures of superconductivity, determined from both susceptibility and resistivity measurements, are shown. This highlights that LBCO near $x$ ${\simeq}$ 1/8 develops an onset $T_{\rm c}$ that exceeds that of LSCO under ambient conditions. The dashed line marks the $x$=0.125 doping level, at which superconductivity is most strongly suppressed in LBCO under ambient conditions. The data for LSCO is taken from Ref. \cite{PhysRevB.40.2254}.}
    \label{fig:enter-label}
\end{figure*}

Next, it is important to address whether the LTT structural phase is suppressed under applied stress. Previously, we reported that in the $x$=0.135 sample the LTT structure is suppressed under stress and replaced by a low-temperature less-orthorhombic phase, as demonstrated by X-ray diffraction measurements.
Here, we performed electrical resistivity measurements on all three compositions, $x$=0.115, 0.125, and 0.135. The resistivity exhibits a hump-like anomaly just before the system enters the SC state, originating from the LTT structural phase transition—specifically, from the pinning of dynamic charge stripes present in the LTO phase as they become statically stabilized upon entering the LTT phase. Thus, resistivity enables us to probe both superconductivity and the LTT structural transition simultaneously. Figures 3a–c show the temperature dependence of the in-plane electrical resistance, normalized to its value at 100 K, measured under various in-plane compressive stresses for the three doping levels $x$=0.115, 0.125, and 0.135, respectively.
We begin with the $x$=0.115 sample. At 0 GPa, the resistance shows a clear increase below $T_{\rm LTT}$=55K, producing a pronounced anomaly before the resistance drops to zero upon entering the superconducting state. The temperature at which zero resistance is reached, Tc,zero, is close to 20 K at 0 GPa. Since we measure in-plane resistance, this reflects the two-dimensional (2D) superconducting transition. The three-dimensional (3D) transition temperature, determined from susceptibility measurements, is lower; thus, the transport measurements probe the higher 2D $T_{\rm c}$, which is consistently higher in all three samples even at zero stress. Upon application of compressive stress, the LTT-related anomaly is gradually suppressed, accompanied by an increase in $T_{\rm c,zero}$. Already at 0.15 GPa, the LTT peak is fully suppressed and $T_{\rm c,zero}$ reaches ${\simeq}$31 K, where it saturates. The same qualitative behavior is observed for the $x$=0.135 sample, with a similar critical pressure of approximately 0.15 GPa required to suppress the LTT anomaly and achieve the optimal $T_{\rm c,zero}$${\simeq}$30K.
In contrast, the $x$=0.125 sample requires a significantly higher critical pressure of about 0.5 GPa—roughly three times larger—to suppress the LTT peak. Moreover, the optimal $T_{\rm c,zero}$ reached under stress is significantly enhanced, attaining 37 K. In addition, the onset of the superconducting transition under the highest applied stress reaches as high as 46 K. Notably, this value slightly exceeds the optimal $T_{\rm c}$ reported for LSCO. This is fully consistent with the susceptibility results presented above, which show that the $x$=0.125 sample—characterized by the most stable stripe state—achieves the highest optimal $T_{\rm c}$ under stress. Figures 3d–f summarize the stress evolution of $T_{\rm c,zero}$ and the LTT-related resistance anomaly ${\Delta}R_{LTT}$ for all three concentrations, revealing an antagonistic relationship between these two quantities. This indicates that suppression of the LTT state is essential for the development of the optimal superconducting state. 


Figure 4 summarizes the main results of our experiments. In Fig. 4a, we present the stress evolution of the spin-stripe ordering temperature $T_{\rm so}$, together with $T_{\rm c,mid}$ and $T_{\rm c,zero}$. The temperature $T_{\rm c,mid}$, extracted from susceptibility measurements, provides a reliable measure of three-dimensional (3D) superconductivity, whereas $T_{\rm c,zero}$, determined from in-plane resistivity, reflects the two-dimensional (2D) superconducting transition. The phase diagram reveals only a slight reduction of $T_{\rm so}$, from 40 K to 33 K, within 0.15 GPa. In contrast, $T_{\rm c,mid}$ increases to 35 K over the same stress range. Meanwhile, $T_{\rm c,zero}$ shows a more gradual increase until it converges with $T_{\rm c,mid}$ at approximately 0.15 GPa, above which a well-defined and sharp superconducting transition is observed in susceptibility. Beyond 0.15 GPa, $T_{\rm c,zero}$ continues to increase slowly, reaching 37 K at 0.56 GPa. Remarkably, the onset of the superconducting transition at 0.56 GPa is as high as ${\simeq}$46 K. Two important conclusions can be drawn from these measurements. First, the $x$=0.125 sample, which hosts the most stable stripe order, develops the highest optimal $T_{\rm c}$ under a relatively small applied stress compared to compositions away from the 1/8 anomaly. Second, the achieved $T_{\rm c,zero}$${\simeq}$37K and onset temperature of ${\simeq}$46 K are comparable to, and even slightly exceed, the optimal $T_{\rm c}$ of LSCO (see Fig. 4d), representing the highest bulk $T_{\rm c}$ reported in the 214 cuprate family. This result is particularly remarkable because the 1/8 stripe-ordered cuprate has long been known for its strongly suppressed superconductivity. Here, however, a modest in-plane stress is sufficient to transform it into a high-$T_{\rm c}$ state. These findings suggest that stripe correlations may, under appropriate conditions, be compatible with—or even favorable to—high-$T_{\rm c}$ superconductivity. Supporting this view, the onset temperature of spin order remains comparable to $T_{\rm c}$. A key prerequisite for the emergence of optimal superconductivity is the suppression of the long-range LTT structural phase, which we identify through the disappearance of the normal-state resistivity peak (see Fig. 4b). This suppression is accompanied by a reduction of the magnetic volume fraction (see Fig. 4b). Importantly, despite this reduction, spin-stripe order remains locally robust under stress, with only modest changes in both the ordered moment size (see Fig. 4c) and the transition temperature. From Fig. 4b, the reduction of the magnetic volume fraction is directly correlated with the suppression of the LTT phase. The enhancement of $T_{\rm c}$ is attributed to the non-LTT fraction, most likely the LTLO phase. At 0.1 GPa, (1-$V_{\rm M}$)  ${\simeq}$ 20${\%}$, which is likely insufficient for percolation. Only when (1-$V_{\rm M}$) approaches 40${\%}$ does $T_{\rm c}$ begin to saturate, indicating that the LTLO phase forms a continuous conducting path. This suggests an inhomogeneous state under stress, with coexisting LTT domains hosting static stripe order and LTLO domains supporting dynamic stripes. Notably, the stress-induced LTLO phase likely differs from its ambient-pressure counterpart, effectively corresponding to a compressed variant along one orthorhombic axis. This distinction is important, as the nominal LTO phase of LSCO has been argued to be LTLO-like, leading to partial stripe pinning and reduced superfluid density and $T_{\rm c}$ near $x$ ${\simeq}$ 0.12, whereas in LBCO the stress-stabilized LTLO phase appears to promote enhanced bulk $T_{\rm c}$.

Overall, uniaxial stress shifts the phase diagram from a regime characterized by strongly suppressed 3D superconductivity—coexisting with the LTT structure and nearly 100${\%}$ static spin-stripe order—toward a high-$T_{\rm c}$ 3D superconducting state with a predominantly orthorhombic structure and long-range spin-stripe order of reduced volume fraction. How can we interpret this evolution? The concept of a pair-density-wave (PDW) state \cite{Fradkin457,lee2023pair,Agterberg231,Tranquada174529,Li067001}  was originally introduced to explain the interlayer frustration \cite{Berg127003} and the suppression of 3D superconducting coherence at 1/8 doping. Within this framework, static spin-stripe order is intertwined with a spatially modulated superconducting order parameter whose sign alternates between neighboring stripes, frustrating Josephson coupling along the $c$-axis and thereby suppressing bulk 3D $T_{\rm c}$. Our results provide direct experimental insight into this picture. ${\mu}$SR measurements show that applied stress reduces the magnetic volume fraction associated with static spin-stripe order. This implies the emergence of non-magnetic regions that do not host static stripe order. These regions are natural candidates for hosting uniform $d$-wave superconductivity. As the fraction and spatial extent of such $d$-wave patches increase, their overlap along the $c$-axis can restore coherent interlayer Josephson coupling, leading to the rapid establishment of 3D superconductivity and the emergence of optimal high-$T_{\rm c}$.
Importantly, the absence of static magnetism in ${\mu}$SR does not necessarily imply the absence of stripe correlations altogether. The non-magnetic patches may still support dynamically fluctuating stripes \cite{Parker2010,Kivelson1201,Huang1161}, which would be invisible to ${\mu}$SR on its characteristic timescale. Thus, stress may not eliminate stripe correlations, but rather convert a fully static stripe configuration into a mixed state comprising static and dynamic stripe regions. In this scenario, uniaxial stress tunes the balance between static and dynamic stripe order. By weakening the static component and enhancing fluctuations, it relieves interlayer frustration, restores $c$-axis coherence, and enables the coexistence of robust 3D superconductivity with stripe correlations. These findings suggest that static stripe order competes with bulk superconductivity primarily through interlayer frustration, whereas stripe fluctuations may instead promote superconducting pairing. In this sense, stripe correlations—when dynamic rather than rigidly pinned—could represent an essential ingredient for high-$T_{\rm c}$ superconductivity in the cuprates.


\section{Conclusion}

In summary, our results reveal that modest in-plane uniaxial stress unlocks a hidden high-$T_{\rm c}$ state in LBCO-1/8 by suppressing the LTT lattice symmetry and consequently converting stripe correlations from static to dynamic. While the stripe-ordering temperature is only moderately affected, the substantial reduction of the static magnetic volume fraction enables the emergence of robust three-dimensional superconductivity. Remarkably, the composition exhibiting the strongest stripe stability and the lowest ambient-pressure $T_{\rm c}$ develops the highest $T_{\rm c}$ under stress, reaching at least 37 K. This finding indicates that the interactions responsible for stripe formation are intimately connected to the superconducting pairing mechanism. Rather than suppressing pairing strength, static stripe order appears to limit global phase coherence, whereas its partial suppression—potentially accompanied by enhanced dynamical stripe correlations—favors optimal superconductivity. These results establish symmetry-selective lattice tuning as a powerful tool to disentangle competing and cooperative orders in cuprates and may provide useful insights into the microscopic origin of high-$T_{\rm c}$ superconductivity.

\section{Methods}

\subsection{Uniaxial stress devices}
For the ${\mu}$SR experiments, we employed a piezoelectric-driven uniaxial stress device designed for operation at cryogenic temperatures, with a sample geometry and size optimized for muon-spin rotation and relaxation measurements. The apparatus is compatible with the Oxford Instruments Heliox $^{3}$He cryostat installed on the general-purpose instrument Dolly at the ${\pi}$E1 beamline of the Paul Scherrer Institute. A detailed description of the uniaxial stress apparatus and the calibration of the force sensor is provided in Refs.~\cite{Hicks2018, Guguchia097005}. A single crystal of La$_{1.875}$Ba$_{0.125}$CuO$_{4}$ (referred to as LBCO-0.125) with dimensions 7${\times}$5.1${\times}$0.3 mm$^{3}$ was used. The crystal was oriented such that the applied stress direction was rotated by 45$^{o}$ with respect to the crystallographic a-axis. The sample was fixed to a Grade-2 Ti sample holder using Stycast 2850FT epoxy. As described in the main text, a pair of coils (100 turns each) was positioned in close proximity to the sample to enable in situ AC-susceptibility (ACS) measurements, allowing us to determine $T_{\rm c}$ of LBCO-0.125 under different stress conditions. To minimize background contributions from the sample holder, hematite absorber pieces were used to mask the regions exposed to the muon beam. Hematite strongly depolarizes the incoming muons, leading to a rapid loss of asymmetry signal. The effective sample area facing the muon beam was therefore 4${\times}$5.1 mm$^{2}$, resulting in approximately 50${\%}$ of the implanted muons stopping within the sample. Based on the calibration, reported previously \cite{Hicks2018,Guguchia097005}  we estimated the compressive force applied to the LBCO-0.125 crystal. The change in the force sensor signal from the zero-force baseline to the maximum applied stress corresponds to 16 ${\mu}$V. We find that a change of 1 ${\mu}$V in the force sensor reading corresponds to an applied force of $\simeq$ 30 N, resulting in a maximum applied force of 480 N. Given the cross sectional area of 1.53 mm$^{2}$, this yields a maximum applied stress of approximately 0.32 GPa.

For the resistivity experiments under uniaxial stress, we employed the FC100 stress cell (Razorbill Instruments) compatible with the Physical Property Measurement System (PPMS, Quantum Design). The single-crystalline La$_{2-x}$Ba$_{x}$CuO$_{4}$ ($x$=0.115, 0.125, 0.135) samples were mounted between two piezoelectrically actuated plates using Stycast epoxy. Measurements were performed using the standard four-probe technique. Razorbill FC100 is a force-controlled stress cell, allowing direct measurement of the force applied on the sample. Knowing the sample width and thickness, the corresponding stress can be determined.\\ 


\subsection{Principles of the \(\mu\)SR Technique}

In a $\mu$SR experiment (see the Supplementary Note 1 and the Supplementary Figure S1), a high-intensity beam of 100\% spin-polarized positive muons is directed into the sample, where the muons thermalize at interstitial lattice sites and act as highly sensitive local magnetic probes. The muons, carrying a momentum of $p_{\mu} = 29$~MeV/c, experience the local magnetic field $B_{\mu}$, causing their spins to precess at the Larmor frequency $\omega_{\mu} = 2\pi \nu_{\mu} = \gamma_{\mu} B_{\mu}$, where $\gamma_{\mu} / (2\pi) = 135.5$~MHz/T is the muon gyromagnetic ratio.

The muon-spin relaxation ($\mu$SR) experiments under zero field (ZF) and weak transverse field (wTF), the latter being applied perpendicular to the initial muon-spin polarization, were performed at the Dolly spectrometer
($\pi$E1 beamline) at the Paul Scherrer Institute, Villigen, Switzerland. 
In this technique, the implanted muons decay with a mean lifetime of $\tau_{\mu} = 2.2~\mu$s, emitting positrons preferentially along the muon spin direction. A set of detectors surrounding the sample records the arrival of muons and the subsequent emission of positrons. The detection process begins when a muon enters the sample, initiating an electronic clock. This clock stops when the decay positron is detected in one of the positron detectors, and the time interval is stored in a histogram. This process is repeated for millions of muon decay events, creating a time-resolved positron count histogram.

In the Dolly instrument, the sample is surrounded by four positron detectors—Forward, Backward, Left, and Right—positioned relative to the muon beam direction. The recorded positron counts in these detectors, $N_{\alpha}(t)$ (where $\alpha = F, B, L, R $), follow an exponential decay due to the muon's finite lifetime. This count distribution also includes a time-dependent polarization function $P(t)$, which encodes the information about the local magnetic environment. The equation governing this decay is given by:

\begin{equation}
N_{\alpha}(t) = N_0 e^{-t/\tau_{\mu}} \left[ 1 + A_0 P(t) n_{\alpha} \right] + N_{\text{bg}}
\end{equation}

where $N_0$ is proportional to the number of recorded events, $A_0$ is the initial asymmetry factor dependent on detector geometry and positron scattering, and $N_{\text{bg}}$ represents the background contribution from uncorrelated events. The initial asymmetry $A_0$ typically ranges between 0.2 and 0.3.

Since positrons are preferentially emitted along the muon spin direction, the signals recorded by the Forward and Backward detectors exhibit oscillations corresponding to the muon precession frequency. To remove the exponential decay component associated with the muon's finite lifetime, a reduced asymmetry function $A(t)$ is introduced:

\begin{equation}
A(t) = \frac{N_{F,L}(t) - N_{B,R}(t)}{N_{F,L}(t) + N_{B,R}(t)} = A_0 P(t)
\end{equation}

where $N_{F,L}(t)$ and $N_{B,R}(t)$ denote the positron counts recorded in the Forward/Left and Backward/Right detectors, respectively. The asymmetry function $A(t)$ and the polarization function $P(t)$ provide direct insight into the static and dynamic properties of the local magnetic environment at the muon site. These functions serve as a powerful tool for investigating the spatial distribution of magnetic fields and their fluctuations in complex materials. For clarity, all ZF and TF \(\mu\)SR asymmetry spectra shown in the main text are plotted after normalization by the initial asymmetry \(A_0\) obtained in the high-temperature paramagnetic state.

\subsection{Analysis of ZF-$\mu$SR data}  

The ZF-$\mu$SR signals shown in were analyzed over the entire temperature range by decomposing the asymmetry signal into contributions from both magnetic and nonmagnetic regions. The zero-field $\mu$SR polarization function is expressed as (see also the Supplementary Note 2):  
\begin{equation}
\begin{split}
P_{\text{ZF}}(t) = & V_{\rm m} \left[ f_{\alpha} e^{-\lambda_T t} J_0(\gamma_{\mu} B_{\text{int}} t) + (1 - f_{\alpha}) e^{-\lambda_L t} \right] \\ + 
& (1 - V_{\rm m}) e^{-\lambda_{\text{nm}} t}.
\end{split}
\end{equation}

Here, $V_{\rm m}$ represents the magnetically ordered volume fraction, while $B_{\text{int}}$ denotes the maximum internal field associated with the Overhauser distribution. The depolarization rates $\lambda_T$ and $\lambda_L$ characterize the transverse and longitudinal relaxation of the magnetic regions, respectively. The parameters $f_{\alpha}$ and $(1 - f_{\alpha})$ correspond to the fractions of the oscillating and non-oscillating components of the magnetic $\mu$SR signal. The function $J_0$ is the zeroth-order Bessel function of the first kind, which describes the incommensurate nature of the spin-density wave and accounts for broad internal field distributions spanning from zero to a maximum value. This behavior is commonly observed in cuprates exhibiting static spin-stripe order~\cite{Guguchia097005, Guguchia214511, Guguchiae2303423120, Klauss4590}. $\lambda_{\text{nm}}$ represents the relaxation rate of the nonmagnetic volume fraction, where spin-stripe order is absent. The complete analysis of $\mu$SR time spectra, including both zero-field (ZF) and transverse-field (TF) measurements, was performed using the open-source $\mu$SR data fitting software \textbf{musrfit}~\cite{Suter2012}.

\subsection{Analysis of Weak-TF \(\mu\)SR Data}

The weak-transverse-field (\(\mu\)SR) asymmetry spectra were analyzed using a model describing the time evolution of the muon-spin polarization (see also the Supplementary Note 3):

\begin{equation}
P_{\text{TF}}(t) = P_{\text{TF}}(0) e^{-\lambda t} \cos(\omega t + \phi),
\end{equation}

where \( P_{\text{TF}}(t) \) represents the muon-spin polarization at time \( t \) after implantation, \( P_{\text{TF}}(0) \) denotes the initial polarization amplitude linked to the paramagnetic volume fraction, and \( \lambda \) is the depolarization rate associated with paramagnetic spin fluctuations and nuclear dipolar interactions. The frequency \( \omega \) corresponds to the Larmor precession, which depends on the applied transverse magnetic field, while \( \phi \) accounts for the phase offset.

For magnetically ordered samples, baseline asymmetry shifts were corrected by allowing \( P_{\text{TF}}(t) \) to be adjusted for each temperature. The magnetically ordered volume fraction at temperature \( T \) was determined using the relation:

\begin{equation}
V_{\rm m}(T) = 1 - P_{\text{TF}}(0,T).
\end{equation}

In the high-temperature paramagnetic phase (\( T > T_{\rm so} \)), where no long-range magnetic order exists, the initial polarization reaches its maximum value, \( P_{\text{TF}}(0,T > T_{\rm so}) = 1 \).

In weak-TF \(\mu\)SR measurements, the signal typically consists of long-lived oscillations from muons precessing in the applied external field, alongside strongly damped oscillations from muons in magnetically ordered regions, where they experience a broad internal field distribution due to the combined influence of applied and internal magnetic fields. ZF-\(\mu\)SR measurements reveal rapid depolarization on a timescale of approximately 0.5~\(\mu\)s. However, in TF-\(\mu\)SR data, variations in the angle between the applied and internal fields create a highly inhomogeneous field distribution in magnetically ordered regions, leading to significant muon-spin dephasing. This, combined with data binning, results in the damped signals from magnetic patches being effectively suppressed in the weak-TF \(\mu\)SR spectra.

\bibliography{main.bib}

\section{Acknowledgments}~
The ${\mu}$SR experiments were carried out at the Swiss Muon Source (S${\mu}$S) Paul Scherrer Institute, Villigen, Switzerland. Z.G. acknowledges support from the Swiss National Science Foundation (SNSF) through SNSF Starting Grant (No. TMSGI2${\_}$211750). Work at Brookhaven is is supported by the Office of Basic Energy Sciences, Materials Sciences and Engineering Division, U.S. Department of Energy under Contract No.\ DE-SC0012704.\\

\section{Author contributions}~
Z.G. conceived, designed and supervised the project. Crystal growth: G.D.G., T.A., I.M., and J.M.. Magnetotransport experiments as a function of uniaxial stress: V.S. and Z.G. with contributions from M.B., A.S., N.A.S.. Preparation for $\mu$SR experiments under stress: Z.G., V.S., S.S.I., M.L.. $\mu$SR experiments under uniaxial stress and corresponding discussions: Z.G., V.S., S.S.I., J.N.G., O.G., P.K., T.S., G.S., R.K., R.S., M.M., M.J., J.C., H.H.K., and H.L.. Data analysis and figure development: Z.G. and V.S.. Writing of the paper: Z.G.. All authors discussed the results, interpretation, and conclusion.\\ 

\section*{Data availability}
The data that support the findings of this study are available from the corresponding authors upon request.\\

\section*{Conflict of Interest}
The authors declare no financial/commercial conflict of interest.\\

\end{document}